# Emission of Linearly Polarized Single Photons from Quantum Dots Contained in Nonpolar, Semipolar, and Polar Sections of Pencil-Like InGaN/GaN Nanowires

Ž. Gačević,[1,*] M. Holmes,[2] E. Chernysheva,[3] M. Müller,[4] A. Torres-Pardo,[5] P. Veit,[4] F. Bertram,[4] J. Christen,[4] J. M. González-Calbet,[5,6] Y. Arakawa,[2,7] E. Calleja,[1] and S. Lazić[3]

[1] *ISOM-ETSIT, Universidad Politécnica de Madrid, Avda. Complutense s/n, 28040 Madrid, Spain*
[2] *Institute for Nano Quantum Information Electronics, The University of Tokyo, 4-6-1 Komaba, Meguro-ku, Tokyo 153-8505, Japan*
[3] *Instituto Nicolás Cabrera and Instituto de Física de Materia Condensada (IFIMAC),*
*Universidad Autónoma de Madrid - Francisco Tomás y Valiente 7, 28049 Madrid, Spain*
[4] *Institute of Experimental Physics, Otto-von-Guericke-University Magdeburg, 39106 Magdeburg, Germany*
[5] *Departamento de Química Inorgánica, Facultad de Químicas, Universidad Complutense CEI Moncloa, 28040 Madrid, Spain*
[6] *ICTS Centro Nacional de Microscopía Electrónica, 28040 Madrid, Spain*
[7] *Institute of Industrial Science, The University of Tokyo, 4-6-1 Komaba, Meguro-ku, Tokyo 153-8505, Japan*

**ABSTRACT**

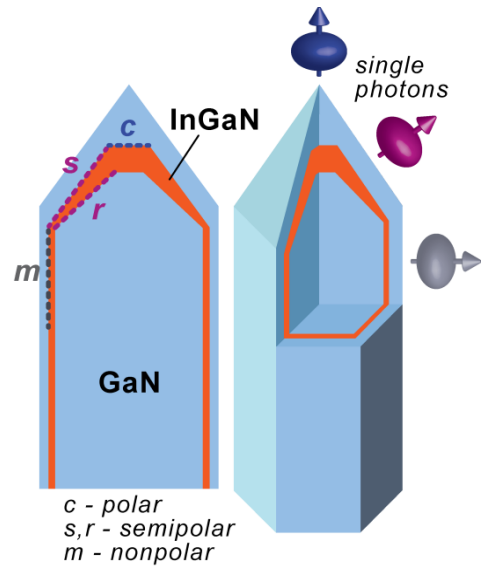

A pencil-like morphology of homoepitaxially grown GaN nanowires is exploited for the fabrication of thin conformal intrawire InGaN nanoshells which host quantum dots in nonpolar, semipolar and polar crystal regions. All three quantum dot types exhibit single photon emission with narrow emission line widths and high degrees of linear optical polarization. The host crystal region strongly affects both single photon wavelength and emission lifetime, reaching subnanosecond time scales for the non- and semipolar quantum dots. Localization sites in the InGaN potential landscape, most likely induced by indium fluctuations across the InGaN nanoshell, are identified as the driving mechanism for the single photon emission. The hereby reported pencil-like InGaN nanoshell is the first single nanostructure able to host all three types of single photon sources and is, thus, a promising building block for tunable quantum light devices integrated into future photonic circuits.

---

[*] Corresponding author: gacevic@isom.upm.es

**INTRODUCTION**

The realization of highly efficient III-nitride classical light emitters has led to a revolution in the solid state lighting market.[1] The key features of III-nitrides which make them a material of choice for classical light emission also qualify them as promising candidates for efficient quantum light emitters that can be operated over a wide spectral range and at high temperatures. Indeed single photon emission has already been demonstrated over range of 280 – 620 nm using III-nitride quantum dots (QDs) of different material composition.[2-15] In addition, the benefits of high band offsets in III-nitride heterostructures have been exploited for strong exciton confinement, thus allowing single photon operation at room temperature[8,9] and even in hot environments (77 ˚C).[13] In other advances, a certain degree of single photon polarization control has been achieved via manipulation of the emitting QD shape,[6,12] and electrically excited single photon emission has also been reported for both columnar and bulk device morphologies.[4,8]

Many of the recent advances in the field of III-nitride-based single photon sources (SPSs) have been achieved using QDs embedded in nanowire (NW) structures.[4,8-11,13-15] Owing to their high crystal quality, GaN NWs can be used to host QDs with superior crystal structure. Furthermore, embedding the QDs into site-controlled NWs allows control over their precise locations. Moreover, the NW heterostructures can be easily transferred from their native to foreign substrates, which facilitates their integration into optoelectronic and photonic circuits.[16]

While in recent years significant progress has been made to extend the III-nitride SPS spectral range, to achieve high-temperature operation and emission polarization control, control over the QD emission lifetime still remains a challenge. In the particular case of III-nitrides with the wurtzite crystal structure, the lack of central symmetry leads to a large internal electric field along the (0001) crystal direction.[17] This (polar) crystal direction is also the preferential one for GaN NW growth,[18,19] resulting in polar "in-wire" nanostructures being the most easily fabricated. The strong internal electric field, induced by the interfaces of these nanostructures, results in an electron-hole spatial separation, which in turn leads to longer emission lifetimes and, accordingly, to a reduction in the maximum possible device operating speed. Faster emission lifetimes are highly desirable in order to boost the operation speeds (repetition rates) of single photon devices. Fast III-nitride-based single photon emission is achievable via a reduction in the electron-hole separation, by either fabricating QDs with smaller size, or reducing the magnitude of the internal electric field.[20-23] Reducing the internal electric field in III-nitride QDs require fabrication in either the zincblende crystal phase[5] (cubic QDs) or QD fabrication on semipolar/nonpolar crystallographic planes in the wurtzite crystal phase (semipolar/nonpolar QDs).[15, 21]

In this article we show that fast single photon emission can be also achieved using III-nitride QDs formed within an InGaN nanoshell grown conformal to the pencil-like morphology of a homoepitaxially grown GaN NW. The NW involves nonpolar *m*-, semipolar *r*- and polar *c*-facets, resulting in the realization of crystal regions of each polarity and hence providing a platform for the fabrication of nonpolar, semipolar, and polar InGaN QDs. We show that single photon emission can be achieved from QDs in each of the three crystal regions, with emission wavelengths from the UV to the visible, and with lifetimes varying from a few hundreds of picoseconds to a few nanoseconds. The pencil-like InGaN nanoshell reported here is the first single nanostructure able to host SPSs with each of the three polarities.

**RESULTS AND DISCUSSION**

The samples were grown by molecular beam epitaxy (MBE) on GaN-on-sapphire templates covered by Ti nanohole masks fabricated through colloidal lithography. The resulting NWs have average diameter of ~180 nm, with the overall diameter variance estimated at ~6%.[24] They are grown in ordered hexagonal matrices with a pitch of ~270 nm. Each NW consists of (i) a pencil-like GaN NW core, (ii) an InGaN nanoshell, and (iii) an external conformal GaN cap (for further details on colloidal lithography and NW growth see Methods section and our previous reports).[25,26] A top-view scanning electron microscopy (SEM) image of an array of structures is shown in Fig. 1a.

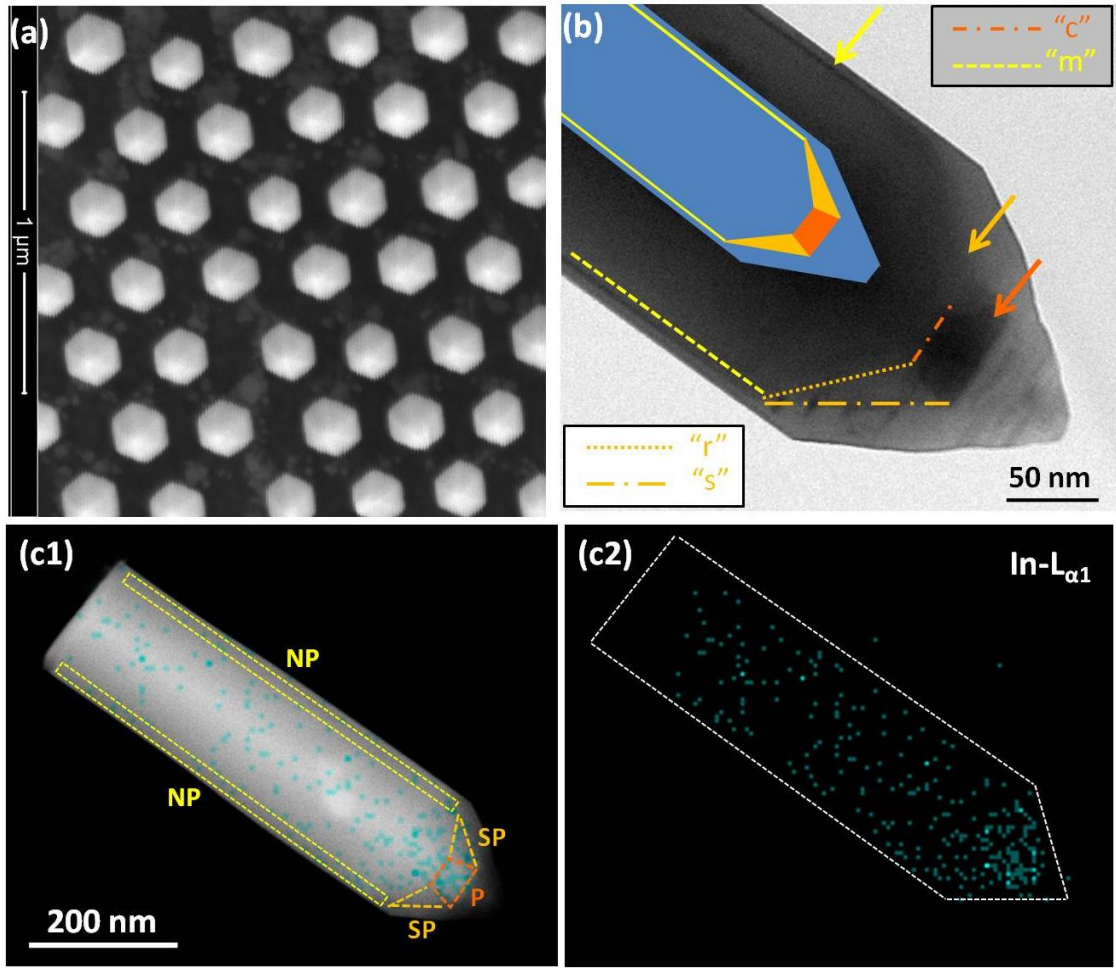

**Figure 1.** (a) Top SEM view of a representative sample area. The hexagonal matrix pitch is ~270 nm and the average NW diameter is ~180 nm. (b) TEM image revealing formation of GaN core, InGaN nanoshell and GaN wrapping as well as crystal planes of the intra-wire interfaces. The nonpolar, semipolar and polar regions of the InGaN nanoshell are designated by arrows, for clarity. (c1) HAADF-STEM image of entire NW combined with spatial distribution of the In-$L\alpha$1 line intensity yields information about In incorporation throughout the studied NW. (c2) The In-$L\alpha$1 line intensity distribution confirms the formation of full conformal InGaN nanoshell, with significantly higher In content in the polar (vs. semipolar/nonpolar) region.

The local structural and chemical characterization of the samples was performed using an aberration-corrected scanning transmission electron microscope (STEM) JEOL-JEMARM200cF (operating voltage: 200 kV, a probe size of ~0.8 nm and a solid semi-angle of 68 - 280 mrad) enabled for energy dispersive X-ray spectroscopy (EDS), using an Oxford INCA-350 detector. Figure 1b shows a TEM image of a single NW confirming the formation of the targeted structure: GaN core, InGaN nanoshell and GaN wrapping (see the schematic inset for clarity). Note that during the InGaN nanoshell growth, the preferential semipolar facet changes from *"r"* (10-12) to *"s"* (10-11), whereas the nonpolar *"m"* (10-10) and polar *"c"* (0001) facets remain unchanged.[26] Panel c1 in figure 1 shows a high angle annular dark field (HAADF) STEM image of the entire NW (diameter/height: 180/600 nm) combined with a map of the spatial distribution of the In-$L\alpha$1 line intensity obtained by EDS. The spatially resolved In-$L\alpha$1 emission is shown in panel c2 for improved clarity (technical details concerning TEM-EDS experiments can be found elsewhere).[26] The mapping confirms the formation of a full conformal InGaN nanoshell around the GaN core. In addition, it reveals a significantly higher In content in the polar nanoshell section with respect to the semipolar/nonpolar ones. This result is attributed to a higher efficiency of In incorporation on polar vs. semipolar/nonpolar crystal planes, as widely reported for nanostructures obtained via MBE.[26-30] The nanoshell thickness varies significantly as a function of crystallographic direction due to different growth rates on polar *m*, semipolar *r/s* and polar *c* planes.[18,19,26] The thinnest and the thickest nanoshell regions are the nonpolar and polar ones, with thicknesses of 1-2 nm and ~30 nm, respectively (cf. Figure 1b).

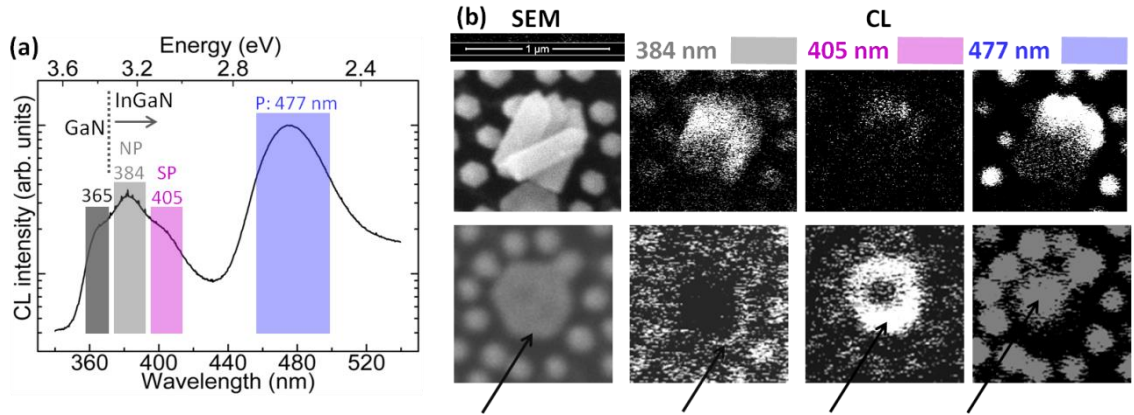

**Figure 2.** Spatial origin of the emission bands as confirmed by RT SEM-CL. (a) CL spectrum of a sample area hosting few hundred NWs reveals four emission bands, one centered at 365 nm, attributed to GaN and three emission bands centered at ~384 nm, ~405 nm and ~477 nm, respectively, attributed to InGaN nanoshells. (b) (up) SEM and SEM-CL images of the sample area containing lying NWs reveal that the ~384 nm emission originates from the NW side walls (i.e. from the non-polar InGaN nanoshell section), whereas ~405 nm and ~477 nm emissions originate from the NW tip. (down) SEM and SEM-CL images of the sample area containing a standing NW with a somewhat bigger diameter (designated by arrow) allows resolving spatial origin of ~405 nm and ~477 nm emission bands, which are attributed to semipolar and polar InGaN nanoshell sections, respectively.

To investigate the emission properties of the InGaN nanoshell, the samples were characterized by room temperature (RT) cathodoluminescence (CL) combined with FEI Inspect F50 SEM (see Methods section). The CL measurements, performed on a few hundred NWs, reveal four emission bands (Figure 2a). Emission centered at ~365 nm, is attributed to GaN near band edge emission; three more emission bands centered at ~384 nm (ultraviolet), ~405 nm (violet) and ~477 nm (blue), respectively, are also recorded. SEM and SEM-CL images of the sample area containing felled NWs (Figure 2b, upper row) reveal that the ~384 nm emission originates from the NW side walls (i.e. from the non-polar InGaN nanoshell section), whereas ~405 nm and ~477 nm emissions originate from the NW tip. SEM and SEM-CL images of the sample area containing an as-grown NW with a larger diameter (Figure 2b, lower row) reaffirms that the ~384 nm emission originates from the nanoshell non-polar section and resolves the spatial origin of ~405 nm and ~477 nm emission bands, attributed to semipolar and polar InGaN nanoshell sections, respectively. Note that this result is in agreement with a detailed study of two-color emitting pencil-like InGaN/GaN nanotips, reported in our separate publication.[26] It is worth noting that in spite of the overall NW diameter estimated at 6% (a certain variance in NW diameters is an intrinsic limitation of the colloidal lithography, see Methods section for further details), NWs with much higher diameters are sporadically observed across the sample surface.

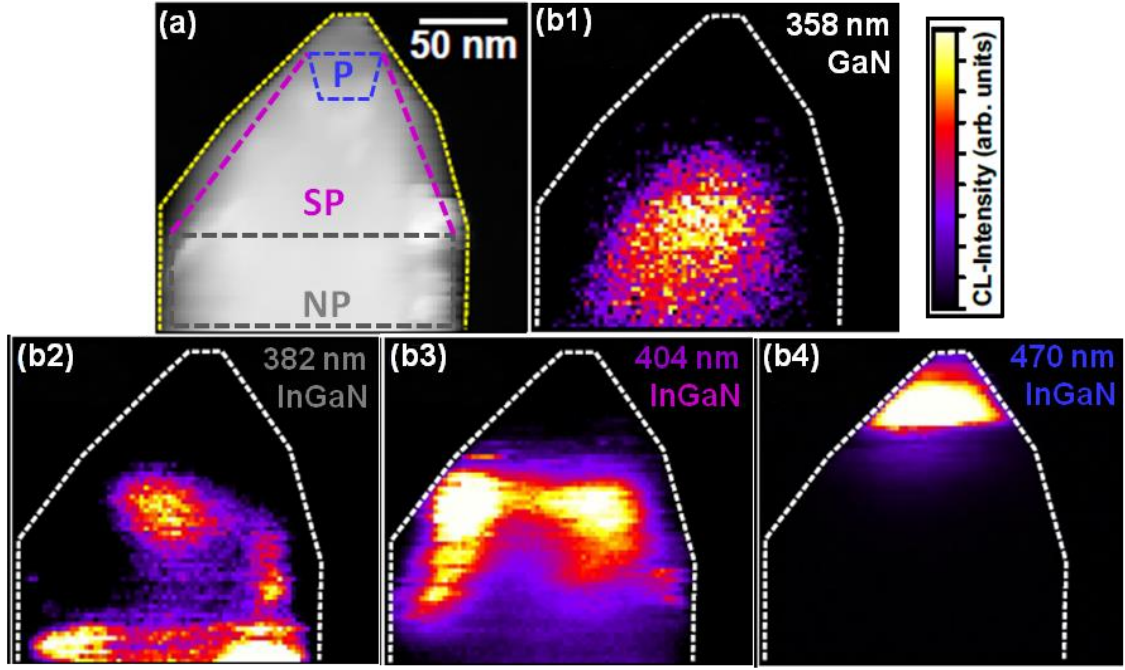

**Figure 3.** The spatial origin of the emissions bands is confirmed by low temperature STEM-CL (T = 16 K) on an individual NW. (a) HAADF image of a NW tip. External dashed lines highlight the external NW facets. Internal dashed lines (guides to the eye) roughly designate the expected positions of the nonpolar, semipolar and polar nanoshell sections. (c1-c4) The ~358 nm emission band originates from the GaN NW core, whereas the ~382 nm (UV), ~404 nm (violet) and ~470 nm (blue) emission bands, originate from the non-polar, semipolar and polar InGaN nanoshell sections, respectively. Strong emission intensity fluctuations are observed across the nonpolar and semipolar nanoshell regions.

Direct correlation of the structural and luminescence properties on a few nanometer scale was achieved using low-temperature CL spectroscopy ($T = 16$ K) in a scanning transmission electron microscope (for technical details see Methods section).[31,32] Figure 3a shows the actual NW tip geometry, as resolved by HAADF-STEM. The HAADF-STEM image clearly resolves the external NW facets (designated with external yellow dashed lines, for clarity) which can be nonpolar, semipolar and polar. In addition, monochromatic CL images taken on the same NW tip area (Figure 3b1-b4) reveal four emission bands in close agreement with RT-CL measurements, featured in Figure 2. The monochromatic CL intensity image of the GaN near bands edge emission at ~358 nm reaffirms that this emission stems from the GaN NW core (cf. Figure 3b1). Furthermore, monochromatic mappings taken at ~382 nm, ~404 nm and ~470 nm, confirm that the UV, violet and blue InGaN emission bands originate from the nanoshell nonpolar, semipolar and polar sections, respectively (cf. Figures 3b2-b4). Apart from strong differences in emission wavelength (from one nanoshell section to another), significant variations in emission intensity within the same section are also directly observed. Figures 3b2 and 3b3 clearly show strong changes in luminescence intensity within both nonpolar and semipolar sections. In addition, local spectra obtained from the InGaN shell region exhibit sharp lines (not shown). These sporadic emission intensity fluctuations together with sharp emission lines are attributed to random In fluctuations across the InGaN nanoshell. The formation of indium rich region within InGaN layer, leading to significant band gap fluctuations and, thus, changing potential landscape, is suggested as origin of exciton localization (at potential minima), as discussed below.

To further assess the optical properties of individual NWs, the NWs were mechanically removed from the native substrate and transferred onto an optically inactive silicon wafer covered with a titanium metal grid (patterned by electron beam lithography). The low density of transferred NWs allows for single NW spectroscopy, whereas the titanium grid containing 3×3 µm$^2$ square apertures facilitates spatial mapping of the NW location. All micro photoluminescence (µPL) measurements in this work were performed on single lying NWs, at ~10 K (experimental details are given in the Methods section).

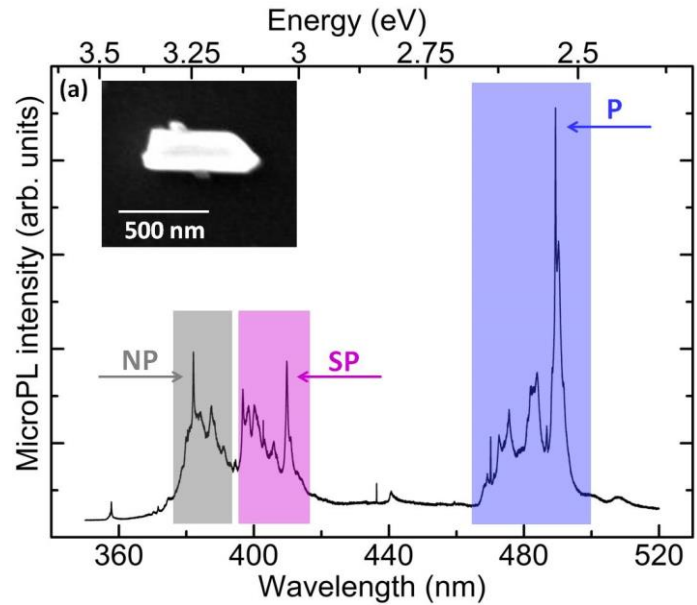

**Figure 4.** The µPL spectrum of a single NW (inset) confirms ultraviolet, violet and blue emission bands originating from the embedded InGaN nanoshell and reveals sharp QD-like emission lines.

The µPL emission spectrum of an isolated single NW (Figure 4) reveals the three InGaN emission bands, in agreement with the CL measurements. In addition, narrow QD-like PL lines, with Lorentzian shape and spectral width $\Delta E < 2$ meV, are observed in all three bands.

To get insight into the nature of the emission (i.e. classical vs. quantum) from these QD-like emission centers, second order coherence $g^{(2)}(\tau)$ measurements (intensity autocorrelation at time delay $\tau$) have been performed. The QD-like emission states were selected in each of the three emission bands, a selection criteria being the spectral width ($\Delta E < 2$ meV) and intensity of their emission (signal-to-total counts ratio $\rho = S/(S + B) > 0.75$, where $S$ and $B$ represent the signal and background counts, respectively). Figure 5 a1-a3 shows the µPL spectra of selected QDs from the nonpolar/semipolar/polar emission bands. The corresponding second order coherence measurements are summarized in Figure 5 b1-b3. All histograms show a pronounced reduction of coincidences at zero time delay (i.e. antibunching), yielding raw $g_m^{(2)}(0)$ values of 0.40 ± 0.09, 0.54 ± 0.06 and 0.33 ± 0.06, for the nonpolar, semipolar and polar QDs, respectively (the subscript *m* stands for "measured"). After correction to account for background contamination[11], the values become $g_{CORR}^{(2)}(0) = 0.09 \pm 0.13$, $0.30 \pm 0.09$ and $0.18 \pm 0.03$ for the nonpolar, semipolar and polar emissions, respectively (note that an additional correction to account for the instrument temporal response was also included for the continuous-wave photon correlation histograms measured from the polar QDs[11]). These values are all well below the 0.5 two-photon threshold, and thus unambiguously show the quantum nature of the emission from the actual localization centers in all three polarity regions.

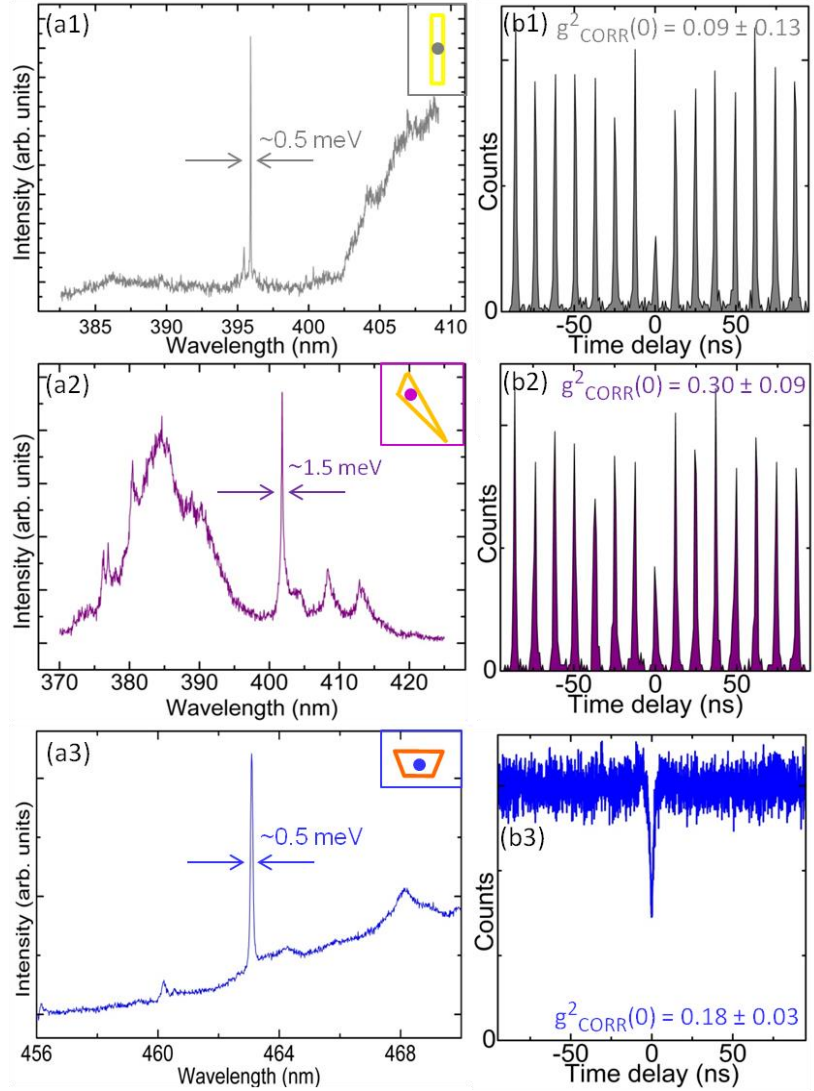

**Figure 5.** (a) Time-averaged μPL and (b) photon correlation measurements of the (1) nonpolar, (2) semipolar and (3) polar QDs. (a1-a3) The emission centers selected for photon correlation measurements exhibit narrow and intense PL lines (linewidth $\Delta E < 2$ meV and signal-to-total counts ratio $\rho > 0.75$). Inset: Sketch depicting the spatial location of the selected QDs within the InGaN nanoshell (see also Fig. 1(b) for clarity). (b1-b3) For all three polarity types, the photon correlation measurements reveal clear antibunching at zero time delay ($\tau = 0$), a signature of single photon emission.

To check consistency with previous reports on III-nitride emitters,[3-8,10-12,14,15] the optical polarization from selected nonpolar, semipolar and polar single photon emission centers has also been measured (see figure 6 for representative measurements). The polarization measurements were performed with respect to a randomly chosen fixed axis. For all three polarity types the degree of linear polarization is greater than 70 %, in agreement with previous reports on InGaN QDs.[6,12,15] This well-known linear polarization is likely due to the anisotropy in the QD confining potential resulting in valence band mixing effects.[33,34]

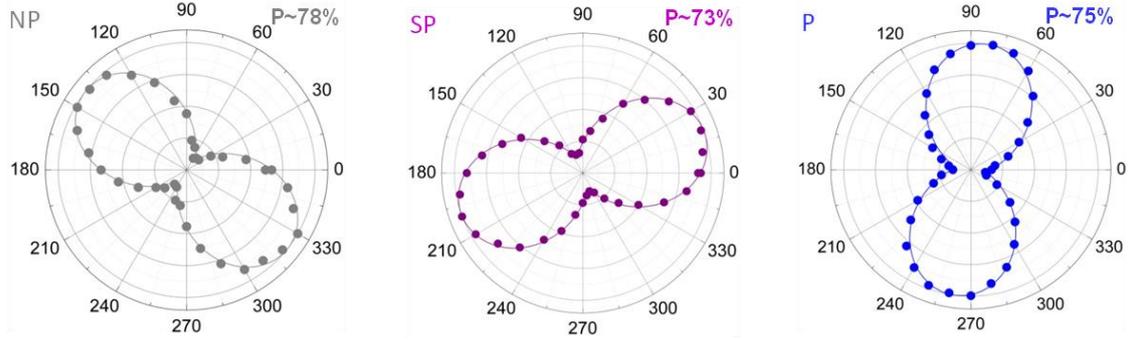

**Figure 6.** Representative polarization-resolved μPL measurements of the single photon emitters from non-polar (left), semipolar (center) and polar (right) reveal high degrees of linear polarization of the emitted photons (>70 %).

Finally, to get insight into their emission lifetimes, time-resolved (TR) μPL measurements were performed on SPSs from each region. Figure 7a shows representative μPL spectra of nonpolar, semipolar and polar InGaN sections, each containing one narrow and intense single photon emission center. The TR-μPL performed on these QD-like centers reveals monoexponential decays with significantly different decay times ($\tau_d$) of ~420, ~860 and ~2250 ps, respectively. More precisely, the TR-μPL performed on several SPSs of each polarity reveals that their typical emission lifetime increases from ~400–600 ps for nonpolar, to ~600–900 ps for semipolar QDs, entering the nanosecond range (> 1.5 ns) for their polar counterparts (Figure 7b). Note that the decay times determined for nonpolar and polar QDs compare well with the values reported by other authors on InGaN QDs fabricated by a different (standard Stransky-Krastanov) approach on nonpolar *a*- ($\tau_d$ ~ 500 ps)[35] and polar *c*- ($\tau_d$ ~ 2 ns)[36] planes.

Previous results demonstrate that a simple technological procedure can be employed for the realization of nonpolar, semipolar and polar InGaN SPSs, all during a single MBE growth process. Experimental measurements reveal that the typical emission linewidths of the three SPS types are ΔE = 0.35 - 2 meV and that they all show similar optical polarization properties (i.e. high degrees of linear polarization P > 70 %). However, it is important to note that the three SPS types differ significantly in their operating spectral range (i.e. photon wavelength) and speed (i.e. emission lifetime). The operating wavelength varies from ultraviolet to violet to blue when the SPS is located in the nonpolar, semipolar and polar InGaN nanoshell sections, respectively (see the insets in Figure 5 a1-a3). Similarly, the corresponding emission lifetime increases from typically ~400–600 to ~600–900 ps, entering the nanosecond range (> 1.5 ns) for polar SPSs (Figure 7b).

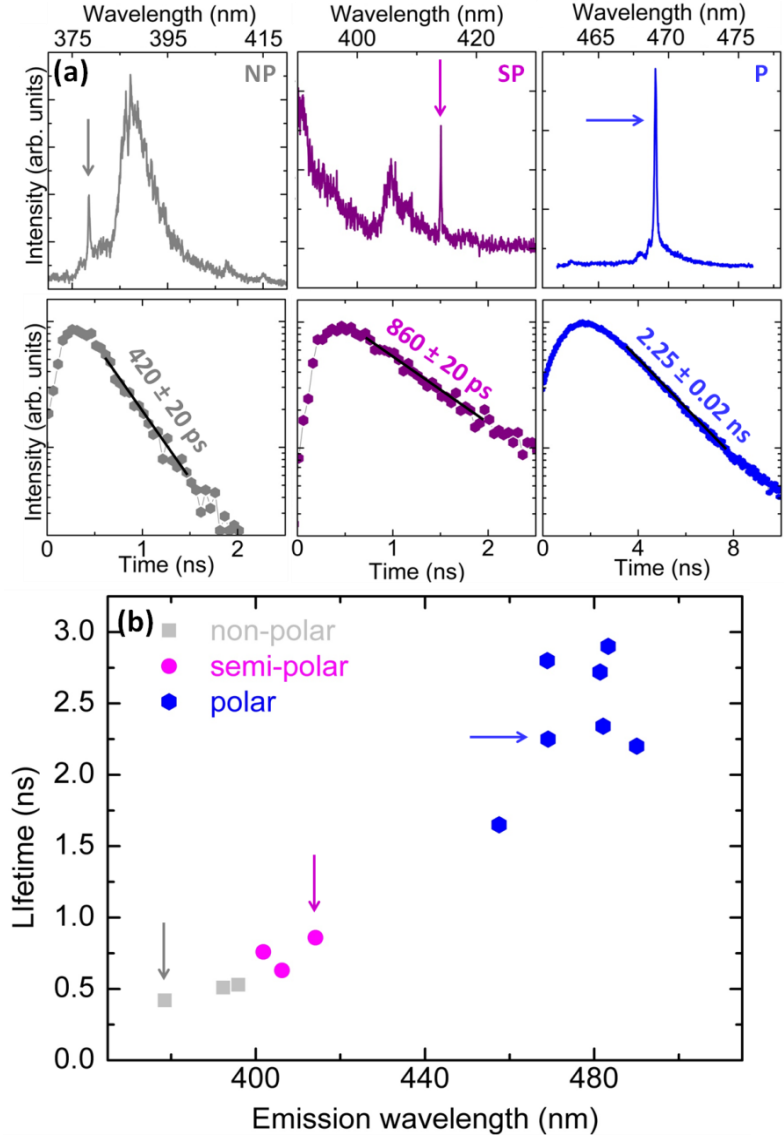

**Figure 7.** (a) Time-resolved μPL decays from nonpolar (NP), semipolar (SP) and polar (P) single-photon emission centers reveal their respective emission lifetimes of ~420, 860 and 2250 ps. (b) Time-resolved μPL performed on several single-photon emission centers with different crystallographic orientations confirms that nonpolar SPSs are typically slightly faster than semipolar ones and significantly faster than their polar counterparts. Arrows indicate SPSs featured in panel (a).

The second order coherence measurements unambiguously confirm the formation of localization sites exhibiting zero-dimensional quantum nature in the InGaN potential landscape of all three nanoshell sections. Note, however, that the nanoshell geometry (in itself) is not the determining factor that leads to the formation of these emission centers. Indeed, it is only along the radial direction of the nonpolar section that the GaN/InGaN/GaN-induced potential variation is on a scale smaller than the exciton Bohr radius (the typically cited value of which is ~3 nm, for III-nitrides).[37] The size of other InGaN nanoshell regions (i.e. polar and semipolar) is too large to expect any significant quantum confinement effect. It is, therefore, clear that the QD-like emission centers are formed within the nanoshell,[11,38,39] most likely due to random fluctuations in the indium content (i.e. In clustering).[40,41] This hypothesis is in a qualitative agreement with strong emission intensity variations observed across InGaN nanoshell sections by TEM-CL (see Figure 3). The high degree of linear polarization, systematically observed for all three SPS types (Figure 6), further (qualitatively) supports this assumption; namely, QDs formed by random In-clustering are expected to have low degree of structural symmetry, resulting in (expectedly) high degrees of linear optical polarization.[6,12,33,34] Also, no preferred crystallographic direction of polarization was observed for QDs from all three InGaN nanoshell regions (not shown), thus indicating that the in-plane anisotropy is not uni-directional, as expected from random asymmetry of these QD structures.

The performance of the three studied SPS types differs significantly in terms of their (characteristic) emission wavelength and emission lifetime. A detailed study concerning the mechanisms responsible for these differences can be found in previous reports, where we have identified the facet-dependent indium incorporation as the dominant mechanism for the change in the emission wavelength.[26] The indium incorporation, which is lowest on nonpolar and highest on polar NW facets, is mainly affected by different properties of the III-nitride dangling bonds of the nonpolar, semipolar and polar crystal planes during the growth process.[26-30] However, it is the facet-dependent internal electric field (the intensity of which is determined by the densities of uncompensated surface charges at nonpolar, semipolar and polar intra-wire interfaces and is lowest in the proximity of nonpolar and highest in the proximity of polar interfaces) that explains the variation in the emission lifetimes between the three regions.[17,26]

**CONCLUSION**

In this article we have shown how the pencil-like morphology of a homoepitaxially grown GaN NW can be exploited for the fabrication of a thin conformal InGaN nanoshell, hosting nonpolar, semipolar and polar SPSs. The three SPS types are characterized with second order photon correlations at zero time delay significantly below the two-photon threshold value ($g_{CORR}^{(2)}(0) < 0.5$), narrow emission linewidths ($\Delta E = 0.35 - 2$ meV) and high degrees of linear optical polarization (P > 70 %). Their properties, however, differ significantly in terms of operating spectral range (ultraviolet, violet, blue) and speed (400–600 ps, ~600–900 ps, > 1.5 ns). These differences are mainly attributed to facet-dependent In content and electric field distribution (respectively) across the hosting InGaN nanoshell. Finally, we identify localization sites in the InGaN potential landscape, most likely induced by random In fluctuations across the nanoshell, as the driving mechanism for SPS formation. These results open the door for the realization of III-nitride quantum light emitting devices with tunable emission lifetimes.

**METHODS**

**Colloidal lithography.** For nanohole masks preparation a commercial colloidal solution, consisting of polyestirene (PS) nanobeads was used. The employed nanobeads have a mean diameter of ~270 nm with ~5% variance. They are sulfate covered i.e. negatively charged. Previous to the lithographic process, the GaN templates were alternately dipped in poly(sodium4-styr-enesolfonate) (PSS) and poly(diallyldimethylammoniumchloride) (PDDA) solutions, which are negatively and positively charged poly-electrolytes, respectively. A tri-layer PSS/PDDA/PSS (~5 nm thick) is formed on the substrate surface to achieve a negatively charged hydrophilic surface. The substrate surface is then covered via spin-coating with the colloidal solution, resulting in a monolayer of PS nanobeads, with a close-packaged arrangement over large areas. Once the nanobeads were arranged on the surface, reactive ion etching oxygen plasma (dissociation power: 20 W, oxygen pressure: 10 mTorr and etching time: 120 sec) was used to reduce the diameter of the beads to ~180 nm and, consequently, to isolate them. After the plasma etching, 7 nm of Ti was evaporated on the surface. Finally, in the lift-off process the colloidal nanobeads are removed, leaving behind a hexagonal 2D array of nanoholes (pitch ~270 nm and diameter ~180 nm) on the Ti layer covering the GaN substrate surface.[42]

**Nanowire growth**. The NWs were grown in a RIBER Compact 21 MBE system, equipped with a radio-frequency plasma nitrogen source and standard Knudsen cells. The growth was performed on commercial GaN-on-sapphire(0001) templates, covered by a ~7 nm thick Ti mask, containing matrices of hexagonally ordered nanoholes previously obtained through colloidal lithography. The growth was realized in a three step-procedure, each step defining one building block: (i) initial pencil-like GaN NW core, (ii) internal InGaN nanoshell and (iii) external conformal GaN wrapping. To achieve growth selectivity, the initial growth step was performed at high temperature ~800 °C, resulting in ordered GaN NWs with pencil-like morphology. To allow In incorporation and the formation of a fully conformal structure, the growth temperature was then decreased to ~600 °C for the growth of both InGaN nanoshell and GaN wrapping.

**SEM-CL.** Direct correlation of the NW optical and morphological properties with a ~100 nm spatial resolution was achieved using room-temperature cathodoluminescence (CL) spectroscopy incorporated within FEI Inspect F50 scanning electron microscope. The emitted CL is collected by a retractable parabolically shaped aluminum mirror, focused onto the entrance slit of a grating monochromator (MonoCL4, Gatan) and detected by a photomultiplier. In parallel, secondary electrons emitted from the sample surface are collected and detected by SEM detection unit. The SEM acceleration voltage is optimized within 5-20 kV range, to provide clear SEM images while preventing luminescence degradation.

**STEM-CL.** Low-temperature high-resolution CL imaging was further performed in a scanning transmission electron microscope to correlate the optical and structural properties of the core-shell nanostructure with a nanoscale spatial resolution.[31,32] The CL detection unit is integrated in an FEI STEM Tecnai F20 equipped with a liquid helium stage. The emitted CL is collected by a retractable parabolically shaped aluminum mirror, focused onto the entrance slit of a grating monochromator (MonoCL4, Gatan) and detected by a liquid nitrogen cooled back-illuminated Si-CCD. Simultaneously to the detection of the CL signal at each position, the electrons that are forward scattered are acquired by an high-angle annular dark-field (HAADF) detector. The STEM acceleration voltage is optimized to 80 kV to minimize sample damage and to prevent luminescence degradation.

**Micro Photoluminescence**. All µPL measurements presented in this work were performed at ~10 K. The spectrally and polarization-resolved time integrated µPL spectra were recorded under a 325 nm continuous-wave He-Cd laser. Measurements were taken using home-built variable temperature microscope setups equipped with single-grating monochromators and charge coupled device (CCD) cameras.

**Photon correlation experiments.** The experiments were performed using a Hanbury-Brown and Twiss (HBT) interferometer[43] placed at an additional monochromator exit port provided with a slit for spectral filtering. Photomultiplier tubes were used in the HBT setup for measuring the nonpolar/semipolar QDs, whereas avalanche photodiodes were employed for the polar ones. In both cases the photon detection events were correlated using commercially available time-correlated single-photon counting electronics. A pulsed frequency tripled Ti:sapphire laser (200 fs pulse-width, 80 MHz repetition rate) at 266 nm was used to excite the nonpolar/semipolar QDs in the HBT experiments, whereas polar QDs were probed with a continuous-wave He-Cd laser 442 nm line. The excitation parameters were chosen so as to achieve optimal conditions (i.e. best signal-to-total counts ratio) for photon correlation measurements. By recording the signal with only one of the HBT photodetectors, the same setups were used for time-resolved (TR) μPL measurements. The aforementioned 266-nm pulsed Ti:sapphire laser 266 nm was used for probing the time-resolved μPL of the nonpolar/semipolar QDs, whereas polar QDs were excited using a pulsed diode laser (<100 ps pulse width, 40 MHz repetition rate) at 405 nm.

**AUTHOR INFORMATION**

Corresponding Author *E-mail: gacevic@isom.upm.es.
The authors declare no competing financial interest.

**ACKNOWLEDGMENTS**

This work was partially supported by Spanish MINECO research grants MAT2015-65120-R, MAT2014-53119-C2-1-R, MAT2011-23068 and MAT2014-58729-JIN, by the Japan Society for the Promotion of Science JSPS KAKENHI projects 15H05700 and 15H06141, by German Research Foundation in the frame of the Research Instrumentation Program INST 272/148-1 and the Collaborative Research Center SFB 787 "Semiconductor Nanophotonics: Materials, Models, Devices." Z.G. acknowledges support from international Erasmus+ project 2015-2-ES01-KA107-022648. E.C. acknowledges the FPI fellowship (MINECO). S.L. acknowledges RyC research grant (MINECO, RyC-2011-09528). E.C. and S.L. acknowledge Prof. Dr. Jose Manuel Calleja for scientific discussions.